\documentclass[aps,prb,reprint,showpacs,twocolumn,superscriptaddress]{revtex4-1}
\usepackage{graphicx}
\usepackage[usenames]{color}
\usepackage{amsmath}
\usepackage{amssymb}
\usepackage{upgreek} 

\newcommand{\qav}[1]{\langle {#1} \rangle}
\newcommand*{\ket}[1]{\left \lvert {#1} \right \rangle}
\newcommand*{\bra}[1]{\left \langle {#1} \right \rvert}

\DeclareMathOperator{\re}{Re}

\begin{document}
\bibliographystyle{apsrev}

\title{Classical-to-quantum crossover in electron on-demand emission}

\author{Vyacheslavs Kashcheyevs}
\email{Corresponding author, email: slava@latnet.lv.}
\affiliation{Faculty of Physics and Mathematics, University of Latvia, Zellu Street 25, LV-1002, Riga, Latvia}
\author{Peter Samuelsson}
\affiliation{Physics Department and NanoLund, Lund University, Box 118, S-221 00, Lund, Sweden}

\begin{abstract}
Emergence of a classical particle trajectory concept from the full quantum description is a key feature of quantum mechanics. Recent progress of solid state on-demand sources has brought single-electron manipulation into the quantum regime, however, the quantum-to-classical crossover remains unprobed. Here we describe theoretically a mechanism for generating  single-electron wave packets by tunneling from a driven localized state, and show how to tune the degree of quantumness. Applying our theory to existing on-demand sources, we demonstrate the feasibility of an experimental investigation of quantum-to-classical crossover for single electrons, and open up yet unexplored potential for few-electron quantum technology devices.
\end{abstract}

\maketitle

\section{Introduction} Single photon on-demand sources have key applications in quantum communication and quantum computation as well as in tests of fundamental properties of quantum mechanics. Large efforts have been directed towards realizing fast and efficient sources, emitting photons in quantum mechanically pure states, one by one \cite{Singphot,Lounis2005}. Single electron on-demand sources in solid state conductors have during the last decade witnessed a similar development  \cite{Feve2007,blumenthal2007a,VK2008b,Pekola2007,fujiwara2008,Hermelin2011,Jehl2013,Dubois2013,Rossi2014}, largely driven by metrological applications of charge quantization \cite{Pekola2013,ROPP2015}. However, in recent years fundamental electron quantum optics experiments with on-demand sources have been performed, such as Hanbury-Brown-Twiss partitioning of single \cite{Feve2012exp,Fletcher2013} and pairs \cite{Ubbelohde2015} of electrons. In particular, indistinguishability and quantum coherence of generated electron excitations have been demonstrated in  seminal experiments via Hong-Ou-Mandel interference \cite{Bocquillon2013,Dubois2013} and Wigner function tomography \cite{Grenier2011,Jullien2014}.

On-demand quantum particle sources also offer a unique possibility to study the emergence of classical properties with unprecedented degree of control. For photons, the conventional description of light in terms of Maxwell equations makes the probability density of coherent states \cite{Glauber1963,Sudarshan1963} a natural classical limit~\cite{Mandel1986}. The quantum-classical transition for photon sources has been experimentally demonstrated~\cite{Bartley2013,Vered2015}.  
For electrons, no classical field limit exists and the appropriate classical notion is that of a point particle on a well-defined trajectory. This is conveniently analysed in terms of the Wigner quasi-probability distribution in phase space \cite{Wigner1932,Case2008}, which approaches a delta-function on the classical trajectory as $\hbar \rightarrow 0$ \cite{Berry1977}.  
Such quantum-classical crossover for individual electrons stands unexplored: experiments with quantum coherent sources have focused so far on fixed-shape wave-functions (either Lorentzian \cite{Dubois2013,Jullien2014} or exponential \cite{Feve2007,Feve2012exp} in time), while measurements on  tunable-barriers emitters \cite{Fletcher2013,Ubbelohde2015} have not yet reached the required temporal and spectral resolution~\cite{Waldie2015,Kataoka2016a,Kataoka2016pss}.

\begin{figure}[bt]
\begin{center} 
\includegraphics[width=0.9\linewidth]{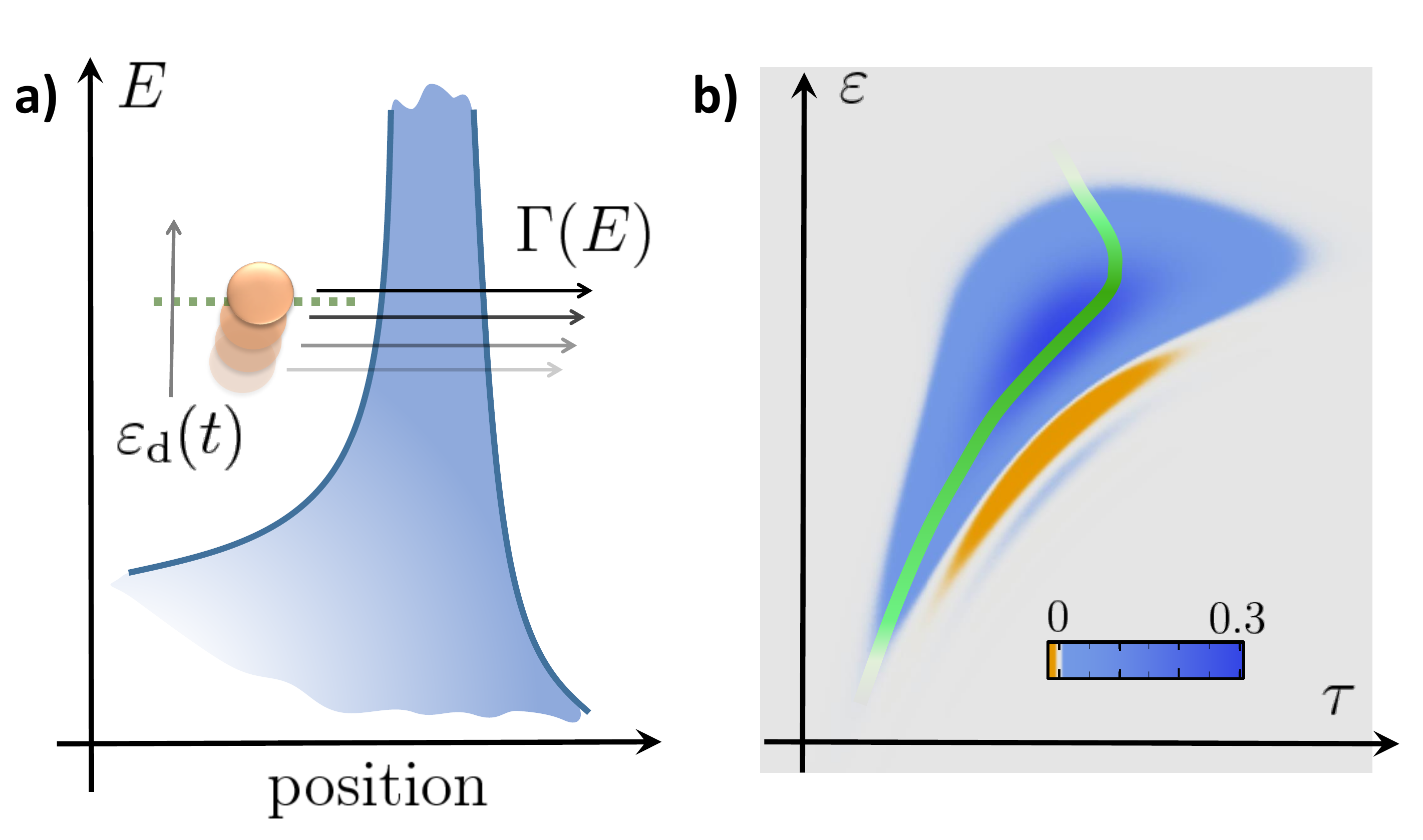}
\end{center}
\vspace{-0.5cm}
\caption{(Color online) (a) Energy-real space schematic of the on-demand source: A single level,
with bare energy  $E=\varepsilon_{\mathrm{d}}(t)$ driven up in time, is
tunnel coupled with an energy-dependent rate  $\Gamma(E)$ to an empty
conduction band. During the drive, an electron initially in the level is
emitted out into the band. (b) Schematic of the energy-time Wigner function $W$
of the free propagating emitted electron with 
signatures of quantum broadening. The green line illustrates the
corresponding classical limit ($\hbar \to 0$) probability density $W_c$, a
delta-function in energy-time space on a guiding trajectory. A specific model used to compute $W$ is discussed in Sec.~\ref{sec:example} below. 
}
\label{fig-system}
\end{figure}

In this paper we show how to design a single-electron source that can be easily tuned between classical and quantum emission regimes.
The design (shown schematically in Fig.~\ref{fig-system}) 
is based on an exact solution for tunneling emission from  a linearly-driven energy level into an empty conduction band, valid for an arbitrary energy dependence of the tunnel coupling density.
We quantify the ``quantumness'' of the source by the spread of the Wigner function of the emitted wave-packet around its guiding trajectory.
To illustrate our results and connect with existing experimental realizations~\cite{Fletcher2013,Ubbelohde2015}, we consider a simple example of energy dependences of the tunnel coupling that allows tuning of the emitted wave-packet from a semiclassical double-exponential \cite{kaestner2010d} via a minimal-uncertainty-product Gaussian~\cite{Ryu2016} to a Lorentzian-in-time {\cite{Keeling2008,Jullien2014} shape by a mere change of the driving rate. Our approach allows for a versatile wave-packet shaping, eliminates the need of phase-matched control signals for emission tuning, and opens new design opportunities for on-demand sources in electron quantum optics.

\section{On-demand tunnelling emission from a driven quantum level}
Initially an electron is localized in a ground state $\ket{d}$ of a sufficiently small quantum dot, separated from an empty band by a high tunnel barrier.
On-demand emission is initiated by driving the quantum dot potential up until the electron tunnels out due to increase of the tunnel coupling with energy. 

\subsection{Classical emission model}
In a classical description of the emission \cite{kaestner2010d,Fletcher2013,Waldie2015}, tunneling out at a time $t_\text{e}$ creates a propagating electron with a well-defined emission energy $E(t)$. For a dot initially populated at time $t_0$, the occupation probability $p_{\mathrm{d}}(t)$ obeys a rate equation $dp_{\mathrm{d}}(t)/dt=-p_{\mathrm{d}}(t)\gamma(t)$ with a time-dependent rate $\gamma(t)$. The resulting distributions of emission times and energies,
\begin{gather} 
\sigma_{\mathrm{c}}(t_\text{e})=-dp_{\mathrm{d}}(t_\text{e})/dt_\text{e} =\gamma(t_\text{e}) \exp \left[-\int^{t_\text{e}}_{t_0} \!\! \gamma(t')\, dt' \right] \, , \label{eq:tdist} \\
\rho_{\mathrm{c}}(E)  =  \sigma_{\mathrm{c}}(t_{\mathrm{e}}(E)) \left|d t_{\mathrm{e}}(E)/dE \right|\, ,
\label{endist}
\end{gather}
are uniquely determined by the externally controlled $\gamma(t)$ and $E(t)$ [via the inverse $t_{\text{e}}(E)$] \cite{Waldie2015}.

For emission into a dispersionless one-dimensional channel, at times $t>t_\text{e}$ the electron propagates away from the dot at constant speed $v_0$, with a simultaneously well defined position $x(t)= v_0(t-t_{\mathrm{e}})$ and momentum $p=E(t_{\mathrm{e}})/v_0$. The Markov tunneling process generates a statistical ensemble along a line in position-momentum space which, at a time $t_{\mathrm{f}} \gg t_{\mathrm{e}}$, provides a direct
imprint of the electron energy dynamics $E(t)$ in the quantum dot before tunneling. Hence it is convenient to use a time $\tau = -x(t_{\mathrm{f}})/v_0 +t_{\mathrm{f}}$ and energy $\epsilon= p v_0$ as the phase space variables characterizing the emitted electron. The corresponding probability density in phase space,
\begin{equation}
W_{\mathrm{c}}(\tau,\epsilon)=\rho_{\mathrm{c}}(\epsilon)\delta(\tau-t_{\mathrm{e}}(\epsilon))=\sigma_{\mathrm{c}}(\tau)\delta(\epsilon-E(\tau)) \, ,
\label{Wigdist}
\end{equation}
is a weighted delta-function along a well-defined trajectory $E(t)$. 

\subsection{Quantum emission model} A relatively slow drive compared to the emission time-scales can be linearized in time as $\dot \varepsilon_{\mathrm d}t$.
The corresponding Hamiltonian of the quantum emission model is
\begin{align} \label{eq:ham}
\mathcal{H}(t) = \dot \varepsilon_{\mathrm d}t \ket{d}\!\bra{d}\!+\!\sum_k E_k\ket{k}\! \bra{k}\!+\!\sum_k V_k\ket{k}\!\bra{d}\!+\!\mathrm{h.c.}
\end{align}
Here $ \ket{d} (\ket{k})$ denotes the level (a band state) and $V_k$ is the 
time-independent amplitude for tunneling between the level and a band state with energy $E_k$. For energies $E_k<E_0=\dot \varepsilon_{\mathrm d}\, t_0$ the amplitude $V_k \rightarrow 0$.  The Shr\"odinger equation for a single-particle state,
\begin{equation}
\ket{\Psi(t)}=c_{\mathrm{d}}(t)\ket{d}+\sum_k \psi_k(t) e^{-i E_k t} \ket{k},
\label{wavefcnexp}
\end{equation}
with initial conditions $c_{\rm d}(t_0)=1$, $\psi_k(t_0)=0$, and Hamiltonian \eqref{eq:ham}, can be reduced to a single
integro-differential equation for the dot amplitude $c_d(t)$ \cite{Basko2016}.
The corresponding conduction band amplitudes are $\psi_k(t)  = -(i/\hbar) V_k \int_{t_0}^{t} e^{i E_k t'/\hbar} c_d(t') \, d t'$. Once the emission is complete, $t>t_{\mathrm{f}}$, $c_d(t) \to 0$ and $(dE_k/dk)^{-1/2}  \psi_k(t) \to \psi(E_k)$, independent of time. The state (\ref{wavefcnexp}) then describes an electron wave packet freely propagating away from the dot, uniquely determined by the emission protocol via $\dot \varepsilon_d$, $V_k$. 
The observable energy and time distributions are $\rho(\epsilon)=|\psi(\epsilon)|^2$ and $\sigma(\tau)=(2 \pi \hbar)^{-1} \lvert \int e^{-i E \tau/\hbar} \psi(E) \, d E \rvert^2$ (shifted to origin by $-t_{\text{f}}$), respectively. Note that $\sigma(\tau)$ can be seen as a distribution of single-electron waiting times \cite{Brandes2008} relative to external trigger (first passage).

Equation \eqref{eq:ham} describes a multi-level Landau-Zener problem, first solved by Demkov and Osherov (DO) \cite{DO} for $|\psi_k|^2$ and later by Macek and Cavagnero (MC)
\cite{Macek1994} for $\psi_k(t)$. Taking the continuum-limit of the DOMC solution, as derived in Appendix~\ref{app:A},
we arrive at 
\begin{equation}
\psi(\epsilon)=  e^{i \phi_k} \sqrt{\frac{\Gamma(\epsilon)}{\dot{\varepsilon}_d}}
\exp \left [ -\frac{i}{ \hbar \dot{\epsilon}_d}  \int^{\epsilon}_{E_0} \left [\Sigma (E) -E \right ] dE
\right ]\, ,
\label{ck}
\end{equation}
where $\Gamma(E)=(2\pi/\hbar)\sum_k|V_k|^2\delta(E_k-E)$ is the tunnel coupling density, $\Sigma(E)=(\hbar/2\pi) \int \Gamma(E)/(E-E'+i 0)\, dE'=\re \Sigma(E)-i\hbar \Gamma(E)/2$ coincides with the retarded self-energy of the quantum dot state $\ket{d}$ due to coupling to the lead, and $\phi_k = \arg V_k +5 \pi/4+{\dot{\varepsilon}}_{\mathrm{d}} t_0^2/\hbar$.

\subsection{Quantum-classical correspondence}
The energy spectrum $\rho(E)$ computed from Eq.~\eqref{ck}  has the same form as $\rho_{\mathrm{c}}(E)$ in Eq.\ (\ref{endist}) if we use a ``naive'' identification of the classical parameters, $E(t)=\dot{\varepsilon}_d t$ and $\gamma(t)=\Gamma\left(E(t) \right)$,  based on the bare (non-renormalized) values for the quantum model. This corresponds to the well-known interpretation of the DO solution \cite{DO} as a sequence of \emph{independent} level-crossing events, each having
a (small) probability of adiabatic transition from $E(t)$ to $E_k$ dictated by the two-level  Landau-Zener formula, $1-\exp [-2\pi |V_k|^2/ (\pi \dot{\varepsilon}_d)]$. The exact distribution of times, $\sigma(t)$, however, involves the energy-dependent level renormalization $\re \Sigma(E)$, and thus converges to $\sigma_c(t)
= \dot{\varepsilon}_d \rho_c(E(t))$ only  in the very restrictive perturbative limit of $\re \Sigma(E) \to 0$.

Here we propose the following non-perturbative definition of the classical trajectory 
$(t^{\star},E^{\star})$ for tunneling emission:
\begin{equation}
E^{\star}=\dot \varepsilon_{\mathrm{d}} t^{\star} + \re \Sigma(E^{\star}) \, .
\label{eq:tstardef}
\end{equation}
A unique function $t^{\star}(E^{\star})$ is always defined by Eq.~\eqref{eq:tstardef}. The inverse, 
$E^{\star}(t^{\star})$ can be interpreted as the fully dressed adiabatic energy of the state $\ket{d}$. Note that 
in case of strong dispersion,  $d \re \Sigma(E)/ d E > 1$,  $E^{\star}(t^{\star})$ becomes multivalued.

To explore the quantum-to-classical crossover, we consider the energy-time Wigner function $W(\tau,\epsilon)=(2 \pi \hbar)^{-1}\!\int \psi^{\ast}(\epsilon-E/2) \, \psi(\epsilon+E/2) e^{-i  E \tau/\hbar}dE $, which can be written as~\cite{Note2}
\begin{multline}
W(\tau,\epsilon) =\frac{1}{2 \pi \hbar}
\int \sqrt{\rho\left(\epsilon\!+\!E/2\right) \rho\left(\epsilon\!-\!E/2\right)} \times \\
\exp\left\{\frac{i}{\hbar} \int_{\epsilon-E/2}^{\epsilon+E/2}\!\!\!\!\! [t^{\star}(E')-\tau ] dE' \right\} dE .
\label{Wigner}
\end{multline}

The relation to the classical limit is elucidated by a saddle-point-type approximation to $W$ via a power series expansion of $\ln \rho(E)$ and $t^{\star}(E)$ around $E =\epsilon$. This gives $W  \! \approx\! \rho(\epsilon) \int e^{\varphi(\omega)} d\omega/(2 \pi) $ with
\begin{align}  \label{eq:saddlepoint}
  \varphi(\omega) &\! =\! i  \omega \left [  t^{\star}(\epsilon)-\tau \right ] - \frac{1}{2} \left [ \omega \, \delta t_Q(\epsilon) \right]^2  + \frac{i}{3} \left [ \omega \, \delta t_{\text{sc}}(\epsilon) \right]^3 \!\! ,
\end{align}
and
\begin{align} 
  \delta t^3_{\text{sc}}(E) &= (\hbar^2/8) \, d^2t^{\star}(E)/dE^2 \, , \\
  \delta t_Q^2(E) & =(\hbar^2/4) \, d^2 [-\ln \rho(E)]/dE^2 \, .
\end{align}
The Wigner function computed from Eq.~\eqref{eq:saddlepoint}} is centered on the trajectory line $\tau=t^{\star}(\epsilon)$ in energy-time space. Explicit analytic evaluation of the corresponding $W(\tau,\epsilon)$ is possible in two limiting cases. In the first case, $|\delta t_{\text{sc}}(\epsilon) | \gg |\delta t_Q(\epsilon)|$, the integral over $E$ in Eq.~\eqref{Wigner} is cut by fast phase oscillations due to $t^{\star}(E)$,
so that the third order in $\omega$ dominates over the second order in Eq.~\eqref{eq:saddlepoint}.
This gives
\begin{equation}
  W_{\mathrm{sc}}(\tau,\epsilon) = \rho(\epsilon) \mathrm{Ai}\left([t^{\star}(\epsilon)-\tau]/\delta t_{\text{sc}}(\epsilon) \right)/\delta t_{\text{sc}}(\epsilon) \, ,
\label{Wignersc}
\end{equation}
where $\mathrm{Ai}$ is the Airy function. The limit of $W \to W_{\text{sc}}$ corresponds to the semi-classical limit defined by Berry for finite quantum systems~\cite{Berry1977}. Equation \eqref{Wignersc} reveals limited quantum fringes on the scale of 
$|\delta t_{\text{sc}}(\epsilon)|$ on the concave side of the classical trajectory~\cite{Berry1977}.

In the other analytic limit of the saddle-point approximation,  $|\delta t_{\text{sc}}(\epsilon) | \ll |\delta t_Q(\epsilon)|$,  
the guiding trajectory is sufficiently straight to be broadened in the temporal direction by the Fourier transform of $\sqrt{\rho(E)}$. Omitting  the term containing
$ \delta t_{\text{sc}}(\epsilon)$ but keeping $ \delta t_Q(\epsilon) $ in Eq.~\eqref{eq:saddlepoint} amounts to a local Gaussian expansion, $\ln \sqrt{\rho (\epsilon\!+\!E/2) \rho (\epsilon\!-\!E/2)}
 \approx \ln \rho (\epsilon) - [E \, \delta t_Q(\epsilon) /\hbar ]^2/2$, with at most linear $t^{\star}(E)$, which gives
\begin{equation}
  W_{\mathrm{G}}(\tau,\epsilon) = \rho(\epsilon)  \frac{1}{ \delta t_{\text{Q}}(\epsilon) \sqrt{ 2 \pi}} \exp \left(-\frac{ 
  [t^{\star}(\epsilon)-\tau]^2}{2 \, \delta t_{\text{Q}}^2(\epsilon) } 
  \right ) \, .
\label{WignerG}
\end{equation}

In the formal limit of $\hbar \to 0$ both 
$\delta t_{\text{sc}}(\epsilon), \delta t_{\text{Q}}(\epsilon)  \to 0$ and the classical expression for the Wigner function, Eq. (\ref{Wigdist}), is recovered with $t_{\text{e}}(E)=t^{\star}(E)$, $E(t)=E^{\star}(t)$, and $\rho_c(E) =\rho(E)$,
thus validating our classical trajectory definition \eqref{eq:tstardef}.  

To quantify the contribution of quantum coherence to the overall spread of an emitted wave-packet, we  express the second moment, $\Delta t$, of the time distribution in terms of energy averages, $\langle \mathcal{F}(E) \rangle \equiv \int \mathcal{F}(E)  \rho(E)\, dE$, as
\begin{equation}
\Delta t^2=
\qav{(t^{\star}-\qav{t^{\star}})^2} + \qav{ \delta t_Q^2 } \, ,
\label{Dt}
\end{equation}
The clear separation  of $t^{\star}$ contribution  motivates us to define a quantumness measure $0\leq \theta \leq 1$ as the fraction of Fourier broadening in the total temporal width, $\theta = \qav{ \delta t_Q^2 }/\Delta t^2$. 

The semiclassical limit \eqref{Wignersc} applies only if $
|\delta t_{\text{sc}}(\epsilon)| \gg | \delta t_{\text{Q}}(\epsilon)|$ which
implies $\theta \ll 1$. The quantum limit, $\theta \to 1$, corresponds to instantaneous emission, $t^{\star}(\epsilon) \approx \text{const}$, with the emission time uncertainty  $\Delta t$ minimized down to the Heisenberg limit for a given (e.g., measured) energy spectrum $\rho(E)$. 
If the latter is  globally Gaussian, then 
$\delta t_{\text{Q}}(\epsilon) =\hbar/(2 \Delta E)=\text{const}$ and
the Heisenberg uncertainty product, $\Delta E \, \Delta t$, in case of fully quantum emission ($\theta = 1$), reaches
the Kennard bound \cite{Kennard1927} of $\hbar /2$. For a Gaussian $\rho(E)$ and at most linear trajectory equation $E^{\star}(t)$, the time distribution $\sigma(\tau)$ is also necessarily Gaussian, with  the measure of quantumness $\theta = (\hbar/2)^2/( \Delta E \, \Delta t)^2=1-\mathrm{R}^2$ reduced by the amount of time-energy correlations. Here $\mathrm{R}$ is the classical Pearson's correlation coefficient which is well-defined for $W_G >0$. 
Note that our quantumness measure $\theta$, which can be maximised by $W\ge 0$ (for example, $W=W_G$), provides a different non-classicality criterion
compared to the Wigner function negativity~\cite{Kenfack2004}, recently adapted
to wavepackets emitted from single particle sources~\cite{Haack2013,Ferraro2013}.

\section{Example: onset of tunnelling density over a finite energy range\label{sec:example}} 
\begin{figure*}
\begin{center}
\includegraphics[width=0.80\linewidth]{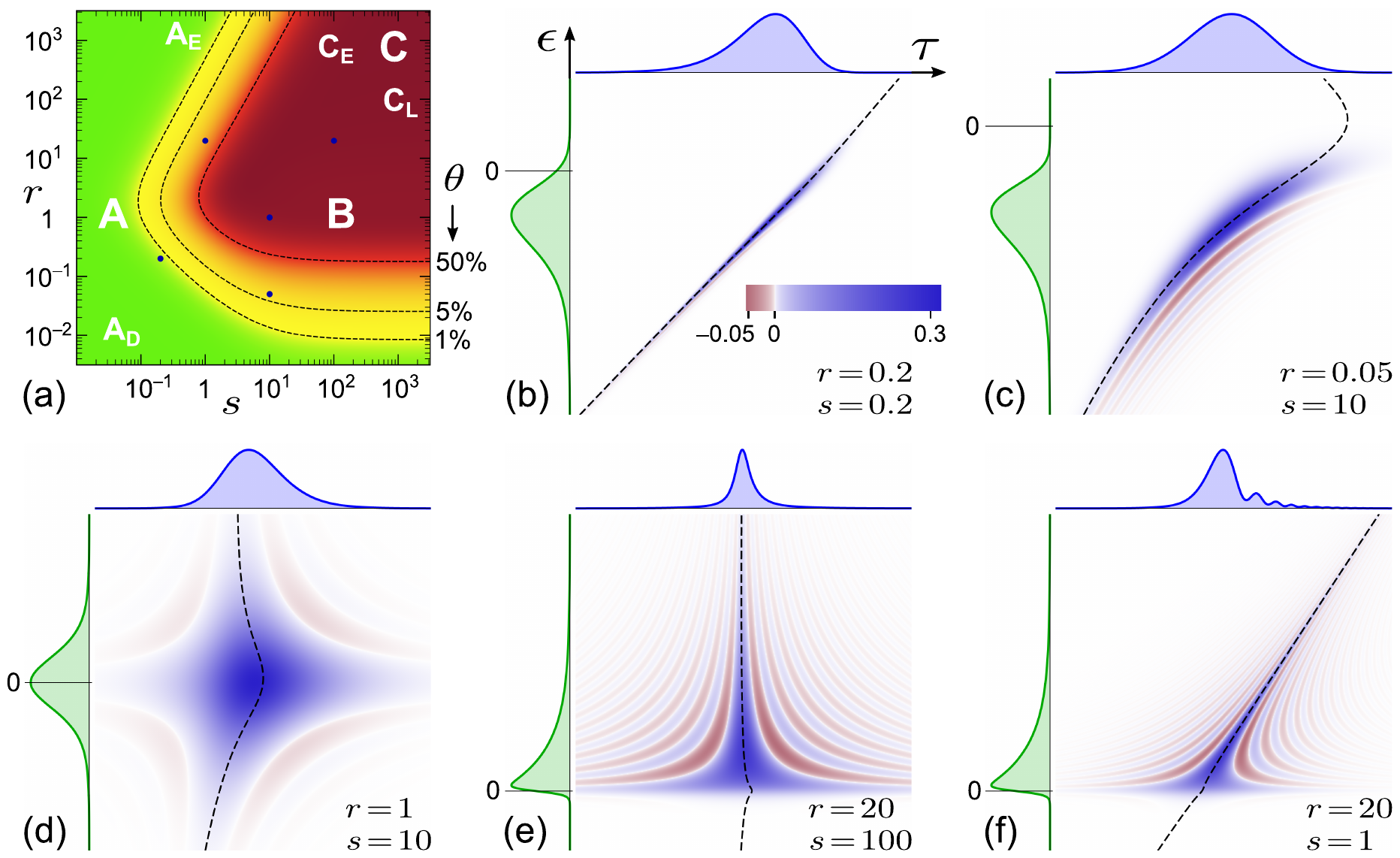} 
\end{center}
\caption{(Color online) (a) Phase diagram  of emission regimes as function of sharpness ($s$) and rapidity ($r$) parameters for the specific tunnel coupling density $\Gamma(E)$ given by Eq.~\eqref{gammaE}. The diagram is colored according to the level of quantumness $\theta(s,r)$ defined by Eq.~\eqref{Dt}, and three levels of $\theta=50\%$ $10\%$, and $1\%$ are indicated for quantitative reference. Regions of 
semiclassical
(A), Heisenberg-limited (B), and  constant-rate 
(C) emission, as well as subregions with double-exponential (A${}_{\text{D}}$), exponential (A${}_{\text{E}}$ and C${}_{\text{E}}$) and Lorentzian (C${}_{\text{L}}$) shape of the time spectrum are indicated; see discussion in the main text. 
(b)-(f) Time-energy Wigner functions $W(\tau,\epsilon)$, in units of $\hbar^{-1}$, for specific values of $s$ and $r$ as indicated. The dashed line traces the guiding trajectory, Eq.~\eqref{eq:tstarspec}. Cf.\ Fig.~\ref{fig-system}(b) computed for $r=0.2$ and $s=10$. The corresponding points in (a) are marked by black circles.  The integrated projections on the upper and the left edge depict the time,
$\sigma(\tau)=\int W(\tau,\epsilon) \, d\epsilon$, and the energy, $\rho(\epsilon)=\int W(\tau,\epsilon) \, d\tau$, distributions, respectively.}
\label{fig:DOfig}
\end{figure*}

We illustrate the  above general results by considering an emitter with the following coupling density,  
\begin{equation}
\Gamma(E)=\Gamma_{\mathrm{0}}\left(1+e^{-E/\Delta_{\mathrm{b}}}\right)^{-1},
\label{gammaE}
\end{equation}
which describes a gradual increase of tunneling from zero to a saturation rate $\Gamma_0$ over a characteristic energy scale $\Delta_b$. 
Emission regimes for the density in Eq.~\eqref{gammaE} are determined by two dimensionless parameters,
level rise ``rapidity'' $r=\dot \varepsilon_{\mathrm{d}}/(\Gamma_{\mathrm{0}}\Delta_{\mathrm{b}})$,
and barrier ``sharpness'' $s=\hbar \Gamma_{\mathrm{0}}/(2\pi \Delta_{\mathrm{b}})$.
Equation~(\ref{ck}) gives the energy distribution
\begin{align}
\rho(E) & =\frac{e^{E/\Delta_b}}{\Delta_{\mathrm{b}}r} \left (1+e^{E/\Delta_{\mathrm{b}}}\right)^{-(1+1/r)} ,
\label{rhoemod}
\end{align}
and Eq.~(\ref{eq:tstardef}) gives the guiding trajectory
\begin{align} 
 t^{\star}  & = \frac{\hbar}{2 \pi \, r \, \Delta_b}\! \left [
 \frac{E}{s \, \Delta_b}  - \re \uppsi \left (\frac{1}{2}\!-\!\frac{i E}{2 \pi \Delta_b}\right ) + \ln \frac{D}{2 \pi \Delta_b} \right ] , \label{eq:tstarspec}
\end{align}
where $\uppsi$ is the digamma function and $D \gg \Delta_b, E$ is the upper cut-off energy of the band.
A complete phase diagram in terms of $\theta(s,r)$, with qualitatively distinct emission regimes marked A, B, and C, and selected examples of $W(\tau,\epsilon)$ are plotted in Fig.~\ref{fig:DOfig}.

The energy distribution \eqref{rhoemod} is controlled by the rapidity $r$ alone, with emission below ($r\!\ll\!1$, cases b, c in Fig.~\ref{fig:DOfig}), above ($r\!\gg\!1$, cases e,f) or at the edge $E\!=\!0$ ($r\!=\!1$, case d). 
The sharpness $s$ controls the shape of the classical trajectory \eqref{eq:tstarspec} and quantum broadening
effects.

For $s \ll 1$, the stationary phase approximation for $W$ applies regardless of $r$,
and the level renormalization is reduced to a constant shift [logarithmic term in Eq.~\eqref{eq:tstarspec}], hence the classical emission model is valid. In Fig.~\ref{fig:DOfig}, this corresponds to region A of the phase diagram (a) and a representative Wigner function (b).
We find  (see Appendix \ref{app:A}) $\Delta E \Delta t/\hbar \sim s^{-1} \max(r, r^{-1}) \gg 1$ as expected. For $r,s \ll 1$ [region $A_{\text{D}}$ in Fig.~\ref{fig:DOfig}(a)]  the energy distribution peaks at $E_{\text{e}} = \Delta_b \ln r$ and both $\rho(E)$ and $\sigma(t)$ are well-approximated by the double exponential form \cite{kaestner2010d}, ie.\  
$\rho(E)=-(d/dE) \exp\left[-e^{(E-E_{\mathrm{e}})/\Delta_{\mathrm{b}}}\right]$.

For $s \gtrsim 1$, a classical-to-quantum crossover can be realized by tuning $r$.
As $r$ is increased towards $1$, the guiding trajectory bends, and the Wigner function develops fringes [see Fig.~\ref{fig:DOfig}(c)], 
in accordance with Eq.~\eqref{Wignersc}.  The classical model \eqref{eq:tdist} and \eqref{endist} is still applicable for $r \lesssim 0.1$ and large $s$, but with parameters, 
renormalised according to Eq.~\eqref{eq:tstarspec}: $E(t)=E^{\star}(t)$ and
$\gamma(t) = \Gamma[E(t)] dE(t)/d(\dot{\varepsilon}_{\text{d}} t)$.

As the rapidity is tuned to $r\!=\!1$, the energy spectrum becomes symmetric. The corresponding arrival time distribution broadening depends on $s$, and the measure of quantumness equals to $\theta(s,r\!=\!1)\!=\!s^2/(1\!+\!1.054\,s^2)$.
For $s \gg 1$, the emitter generates quasi-Gaussian, Heisenberg-limited wave-packets with $\Delta E \, \Delta t  \to \sqrt{(4+\pi^2)/48} \approx  0.538 \hbar$, which is close to the tightest possible simultaneous localization in time and energy, see Fig.~\ref{fig:DOfig}(d) and region B in Fig.~\ref{fig:DOfig}(a). The corresponding width $\Delta E=\pi \sqrt{3} \Delta_b \gg \hbar\Gamma_0$ is set by the energy-dependence (but not the absolute value) of $\Gamma(E)$.

Finally, the limit of $\Delta_b \to 0$ [$s$, $r\!\to\!\infty$, region C in Fig.~\ref{fig:DOfig}(a)] corresponds to a sudden onset of emission at a constant rate $\Gamma_0$ and is equivalent to the zero-temperature limit of a linearly driven small mesoscopic capacitor~\cite{Keeling2008}. The energy spectrum is a simple exponential with $\Delta E\!=\!\dot \varepsilon_{\mathrm{d}}/\Gamma_{\mathrm{0}}$ while the time distribution crosses over from a Lorentzian at $1 \ll r \ll s$ [see Fig.~\ref{fig:DOfig}(e) and region $C_{\text{L}}$ in Fig.~\ref{fig:DOfig}(a)]  via an oscillating regime at $s \sim r$ [similar to Fig.~\ref{fig:DOfig}(f)] to an exponential at $1 \ll s \ll r$ [region $C_{\text{E}}$ in Fig.~\ref{fig:DOfig}(a)], in exact accord with Ref.~\onlinecite{Keeling2008}. Although the overall shape of both $\sigma(t)$ and $\rho(E)$ is exponential for $r \gg s \gg 1$ and thus consistent with the classical relation \eqref{endist}, the quantumness measure $\theta$ drops from $1$ to $0$ only at $r \gtrsim s^2$,  see the boundary between regions $A_{\text{E}}$ and $C_{\text{E}}$ in Fig.~\ref{fig:DOfig}(a). This is because the quantum contribution to the second moment $\Delta t$  is sensitive to the tails of $\sigma(t)$, and the latter  are broadened by a small but non-zero $\Delta_b$ (such a regime is beyond the dispersionless model of Ref.~\onlinecite{Keeling2008}). The product of uncertainties always remains large for large rapidities: $\Delta E \Delta t / \hbar \sim \max(\sqrt{r},r/s) \gg 1$ for $r \gg 1$.
Note that in contrast to temperature in Fermi-sea-triggered emitters \cite{Feve2007,Keeling2008,Bocquillon2014}, finite $\Delta_b$ allows for \emph{coherent} shaping of wave-packets, exemplified by regime $B$ discussed above.

\section{Feasibility and generalizations}
The single-particle approach adopted in our model for the electron emission is justified for experimental realisations where electron is emitted well above the Fermi energy~\cite{Fletcher2013,Ubbelohde2015,vanZanten2016}. The strong coupling regime (ie.\ essential renormalization and non-classical emission) is reached via  the competition of the tunnel coupling strength with the characteristic scale for its variability in energy (e.g., $\Delta_b$ in the example of Sec.~\ref{sec:example}). Both may still be significantly smaller than other energies scales relevant for the localized state physics, such as level spacing, Coulomb charging energy, Kondo scale, superconducting gap etc.

Experiments with single-electron emission from tunable-barrier quantum dots coupled to ballistic edge channels in GaAs~\cite{kaestner2010d,Fletcher2013,Ubbelohde2015,Kataoka2016a} have recently demonstrated \cite{Waldie2015} $\Delta E \, \Delta t \leq (1.0\,\text{meV}) (2.7\,\text{ps})  \sim  4.2 \hbar$ with contributions to $\Delta t$ due to classical time-energy correlations \cite{Kataoka2016pss}, putting the quantum limit within reach. In addition to lithographic and electrostatic confinement,
individual impurities~\cite{Roche2013} or superconductors \cite{vanZanten2016,Basko2016} may be used to tailor $\Gamma(E)$. 

Our general analysis of quantumness in terms of Eqs.~\eqref{ck}--\eqref{Dt}
does not rely on the explicit DO solution \eqref{ck}, and applies to an arbitrary coherent wave-packet $\psi(E)=\sqrt{\rho(E)} \exp [ i \int^E t^{\star}(E') d E'/\hbar]$ with $\rho(E)$ and $t^{\star}(E)$ derived from a microscopic quantum model, appropriate for a particular barrier and protocol design, e.g., 
a tuneable barrier with known $\epsilon(t)$ and $V_k(t)$ \cite{Battista2012,Gurvitz2015,Basko2016} or a real-space model beyond single-level approximation~\cite{Ryu2016}.
Generalization of the quantumness criterion for on-demand single-particle excitations to mixed states~\cite{Ferraro2013} and  many-body systems~\cite{Calzona2016,Litinski2016} is a promising avenue for further research.

\section{Conclusions}
We propose the use of a statically structured tunnel coupling density to control the time and energy distribution of coherent electrons emitted on demand. Using an exact non-Markovian solution for spontaneous electron emission from an initially localized pure state into a one-dimensional ballistic channel, we have theoretically demonstrated the feasibility of a crossover from semiclassical to quantum-limited wavepacket emission which can be realized as a function of the driving rate alone using a suitable $\Gamma(E)$.\!\!
This opens new possibilities for engineering solid-state electron wavepackets that have a broad application potential
from basic studies of entanglement in solid state \cite{SplettstoesserInterf2009,Sherkunov2012,Bocquillon2013,Ubbelohde2015}  to electronic quantum technology, such as ultafast voltage sampling \cite{Johnson2016}.

\begin{acknowledgments}
We acknowledge discussions with G.~F{\`{e}}ve, J. Rammer, J. Erdmanis, E. Locane, and F.~Vi\~{n}as. This work has been supported by the Latvian Council of Science (V.K., grant no.\ 146/2012), Swedish Science Foundation (P.S) and NanoLund (V.K.).
\end{acknowledgments}

\appendix

\section{Continuous limit of DO-MC solution\label{app:A}}

Here we derive Eq.~\eqref{ck} for the asymptotic amplitude $\psi(E_k) = (dE_k/dk)^{-1/2} \lim\limits_{t\to \infty}  \psi_k(t) $ for electron emission from a localized state $\ket{d}$ at $t_0 \to -\infty$ into of a normalized quasi-continuous scattering state $\ket{k}$ with energy $E_k$, as defined by Eqs.~\eqref{eq:ham} and \eqref{wavefcnexp}.

The wave function $\ket{\Psi(t)}$ defined by Eq.~\eqref{wavefcnexp} in the main text 
can be written in terms of a time evolution operator $U(t,t_0)$, with $t \geq t_0$ which has been computed by MC for a discrete set of levels with arbitrary $V_k$ and $E_k$ (multi-level Landau-Zener problem).  
We are interested in the solution of the initial value problem with $\ket{\Psi(t_0)}=\ket{d}$, 
so the required amplitude is
\begin{equation} 
c_k(t) \equiv \psi_k(t) e^{-i E_k t}  =\bra{k}U(t,t_0)\ket{d} \, .
\label{ck1}
\end{equation}
This quantity is given by Eq.~(47) of MC paper,
which in our notation reads
\begin{widetext}
\begin{equation}
c_k(t)=\frac{V_k}{2\pi\hbar\dot\varepsilon_{\mathrm{d}}}\int_{-\infty}^{\infty} \left[\frac{e^{i\phi(E)-iEt/\hbar}}{E-E_k+i0}  \int_{-\infty}^{E} e^{-i\phi(\epsilon)+i\epsilon t_0/\hbar} d\epsilon  \right] dE \, ,
\label{ckexp}
\end{equation}
\end{widetext}
where the complex phase function is defined by 
\begin{equation} 
\phi(E)=\frac{1}{\hbar \dot\varepsilon_{\mathrm{d}}} \int^{E}_{E_\text{min}} \left[E'-\sum_q \frac{|V_q|^2}{E'-E_q+i0}\right] \, dE' \, .
\end{equation}

Taking the continuum limit, the sum over $q$ in turns to an integral over $E$ and  $\phi(E)$  becomes 
\begin{equation} \label{eq:phi}
\phi(E)=\frac{1}{\hbar \dot\varepsilon_{\mathrm{d}}} \int^{E}_{E_{\text{min}}} \left[E'-\Sigma(E')\right] \, dE' \, , 
\end{equation}
where $\Sigma(E)$ is the retarded self energy defined in the main text.
Note that the value of $c_k(t)$ is independent of the lower limit $E_\text{min}$ of the energy integral in  \eqref{eq:phi} as long as $E_\text{min}$ 
is smaller than any relevant $E_k$, i.e., $\Gamma(E) =0$ and $\phi(E)$ is real for $E < E_{\text{min}}$. Both $\Sigma(E)$ and 
$\phi(E)$ are holomorphic in the upper half-plane of complex $E$.

In the limit of $t_0 \rightarrow -\infty$ the integral in the bracket in Eq.~(\ref{ckexp}) can be evaluated exatly by the method of  stationary phase. The stationary point energy $E_0$ is the solution to $d[\re \phi(E) -E t_0/\hbar]/dE=0$ which is given by $t_0=[E_0-\re\Sigma(E_0)]/\dot \varepsilon_{\mathrm{b}}=t^{\star}(E_0)$. Taking into account that for $t_0 \rightarrow -\infty$ we have $\Sigma(E_0) \rightarrow 0$  and hence $E_0\rightarrow \dot\varepsilon_{\mathrm{d}} t_0$, the saddle point evaluation gives
\begin{equation}
\int_{-\infty}^{E} \! e^{-i\phi(\epsilon)+i\epsilon t_0/\hbar}d\epsilon
=\sqrt{\frac{2\pi}{|\phi''(E_0)|}}e^{i\pi/4}e^{-i\phi(E_0)+iE_0t_0/\hbar} \, ,\end{equation}
where $\phi''(E_0)=d^2\phi(E)/dE^2|_{E=E_0}\to 1/(\hbar \dot \varepsilon_{\mathrm{d}})$. We thus have the Heisenberg picture solution for the asymptotic initial condition $\ket{\Psi(t_{0}\!\to\!-\infty)} =\ket{d}$,
\begin{equation} \label{eq:intermediate}
c_k(t)=\frac{V_k e^{i\pi/4}}{\sqrt{2\pi\hbar \dot\varepsilon_{\mathrm{d}}}} e^{-i\phi(E_0)+iE_0t_0/\hbar}\int_{-\infty}^{\infty}  \frac{e^{i\phi(E)-iEt/\hbar}dE }{E-E_k+i0} \, .
\end{equation}

In the limit of $t\rightarrow \infty$ we can perform the remaining integral by contour integration. We shift the contour down into  the lower half of the complex energy plane, so that it runs  parallel to the real axis with a small negative imaginary part $-i|\eta|$ to the integration variable $E$. The value of $\eta$ is chosen such that the pole at $E=E_k-i0$ is enclosed but none of the poles of  $e^{i\phi(E)}$ are. For $t\rightarrow \infty$, the value of the integrand along the shifted contour is exponentially suppressed as $e^{-|\eta|t}$ and can be neglected. The  integral in Eq.~\eqref{eq:intermediate} thus evaluates to  $-2\pi e^{i\phi(E_k)-iE_kt/\hbar}$, and the expression for $c_k(t)$ becomes
\begin{align}
c_k(t)&=-V_k\sqrt{\frac{2\pi}{\hbar \dot\varepsilon_{\mathrm{d}}}}e^{i\pi/4}e^{i[\phi(E_k)-\phi(E_0)]-i(E_kt-E_0t_0)/\hbar} \, .
\label{ck1text}
\end{align}
Taking into account the definitions \eqref{eq:phi} and \eqref{ck1}, one recognizes \eqref{ck1text} as Eq.~\eqref{ck} of the main text. Note that up to the initial phase factor, the amplitude $\psi_k$ is the scattering matrix element for $\ket{d} \to \ket{k}$ transition with the time-dependent scattering potential defined by $V_k$ and $\dot{\varepsilon}_{\mathrm{d}} (t-t_0)$. 

\section{Analytic results for the specific emission model}

Here we provide a derivation of the time and energy distribution parameters for the barrier with energy dependent rate $\Gamma(E)=\Gamma_{\mathrm{0}}[1+\exp(E/\Delta_{\mathrm{b}})]^{-1}$ presented in Eq.~\eqref{gammaE} of the main text. In particular, we derive explicit results for different limits of the rapidity $r=\dot \varepsilon_{\mathrm{d}}/(\Gamma_{\mathrm{0}}\Delta_{\mathrm{b}})$ and the sharpness $s=\hbar\Gamma_{\mathrm{0}}/(2\pi \Delta_{\mathrm{b}})$. 

The self-energy function $\Sigma(z)$ for this special case is
\begin{align}
 & \Sigma(\omega \pm i 0) = \frac{\hbar}{2 \pi} \int_{-\infty}^{D} \frac{\Gamma(E)}{\omega-E \pm i 0} d E \nonumber \\
  & =  s \Delta_b \left [ \uppsi\left (\frac{1}{2} \mp i\frac{\omega}{2 \pi \Delta_b} \right )
    \mp  i\frac{\pi}{2}  - \ln \frac{D }{2 \pi \Delta_b} \right ] \, , \label{eq:ourSigma}
\end{align}
with exponential accuracy for $D-\omega \gg \Delta_b$. Equation \eqref{eq:ourSigma} leads directly to the trajectory equation given by 
~\eqref{eq:tstarspec} of the main text.

The emission energy distribution $\rho(E)$ is given by 
computing $|\psi(E)|^2$ to which only the imaginary part of $\Sigma(E)$ contributes. The specific form  \eqref{gammaE} of the latter can be integrated analytically which yields Eq.~\eqref{rhoemod}. The first two central moments of the energy distribution 
can also be computed explicitly, 
\begin{align}
\langle E \rangle & =-\Delta_{\mathrm{b}} \label{delE0}
\left[\gamma_{\mathrm{Euler}}+\uppsi(1/r)\right] \, , \\
\Delta E & =\Delta_{\mathrm{b}}\sqrt{\frac{\pi^2}{6}+\uppsi_1(1/r)} \, ,
\label{delE}
\end{align}
where $\gamma_{\mathrm{Euler}}= 0.577\ldots$ is the Euler constant and $\uppsi_1(z)=d\uppsi(z)/dz$ is the trigamma function.  

For computing the time uncertainty $\Delta t$, we note that the Fourier transform of $\sigma(\tau)=\int W(\tau,\epsilon) \, d\epsilon$  is by definition the generating function for the moments,
$\langle t^{n} \rangle_t$, where $n=0, 1,2 \ldots$ and 
 $\langle \cdots \rangle_t=\int \cdots \sigma(t) \, dt $. 
With this argument, the general expression \eqref{Wigner} for the Wigner function leads
to Eq.~\eqref{Dt} where  
$(\Delta t)^2\equiv \langle t^2\rangle_t-\langle t\rangle_t^2=(\Delta t^{\star})^2+(\Delta t_{\mathrm{Q}})^2$ is expressed via 
$(\Delta t^{\star})^2=\qav{[t^{\star}(E)-\qav{t^{\star}(E)}]^2}  $ and $(\Delta t_{\mathrm{Q}})^2 =
\qav{ \delta t_Q(E)^2 }$.

For the specific $\Gamma(E)$ given by Eq.~\eqref{gammaE}, 
the quantum contribution to time-broadening can be evaluated analytically for any $r$,
\begin{equation}
\Delta t_{\mathrm {Q}}=\frac{\hbar}{2}\sqrt{-\left\langle \frac{d^2 \ln\rho(E)}{dE^2}\right\rangle}=\frac{\hbar}{2\Delta_{\mathrm{b}}}\sqrt{\frac{1}{2r+1}} \, .
\label{tgwidth}
\end{equation}

The semiclassical contribution $\Delta t^{\star}$ involves integrals $\langle (\re\Sigma(E)-E)^n \rangle_E$ for $n=1,2$ which 
we could not perform analytically for arbitrary $r,s$.  Asymptotic limits of the moments of time and energy distributions for $r \to 0, 1$, and $\infty$  are summarized  in Table~\ref{tab:moments}. These results have been used in the description of different emission regimes in the main text.

\begin{table*}
\caption{\label{tab:moments}Time-energy uncertainty budget in different limits for the specific emission  model defined by Eq.~\eqref{gammaE}.}
\begin{ruledtabular}
\begin{tabular}{c|c|c|c}
Quantity &  $r \ll 1$  & $r \gg 1$ &  $r=1$  \\ \hline
$\Delta_b \rho(x\!=\!E/\Delta_b)$    
& 
   $ r^{-1} \exp\left[-r^{-1} e^{x}+x\right] $  
&  $ \displaystyle r^{-1} \Theta(x)  e^{-x/r} $
& $ \displaystyle \left[ 2 \cosh(x/2) \right]^{-2} $\\
$\qav{E}/\Delta_b$    
&  $-\gamma_{\mathrm{Euler}} - |\ln r | $
&  $r$
&  0
\\ 
$\Delta E / \Delta_b$
& ${\pi }/{\sqrt{6}} $
& $r$
& ${\pi}/{\sqrt{3}} $ 
\\
$\Delta t_Q /( \hbar \Delta_b^{-1}) $
&$ 1/2 $
& $(8 r)^{-1/2}$
& $ 1/ (2 \sqrt{3})$
\\
$\Delta t^{\star} /( \hbar \Delta_b^{-1})$
& $ \displaystyle \left [(r s)^{-1} + (r | \ln r|)^{-1} \right ]/(2 \sqrt{6}) $
& $(2 \pi)^{-1} \left [(r^{-1}-s^{-1})^2 + (\pi^2/6\!-\!1) r^{-2} \right]^{1/2}$
& $\left (a_2-a_1^2+s^{-2} /12 \right )^{1/2}$ \footnotemark[1]  
\\
$\Delta E \, \Delta t / \hbar$
& $ \displaystyle  \pi \left  [(r s)^{-1} + (r | \ln r|)^{-1} \right ]/12 $
& $ \displaystyle  \left \{ 
\frac{r}{8}\left [ 1+\frac{2}{\pi^2 s}\left(\frac{r}{s}-2\right)\right ]
\right \}^{1/2} $
& $(\pi/6)  \displaystyle  \left [\frac{3}{4} + \frac{3}{\pi^2}+s^{-2} \right ]^{1/2} $ 
\\
\end{tabular}
\end{ruledtabular}
\footnotetext[1]{Here $a_n\equiv
\int \left[ 2 \cosh(x/2) \right]^{-2}  \left(\frac{1}{2\pi} \re \left\{\uppsi\left(\frac{1}{2}+i \frac{x}{2\pi}\right)\right\}\right)^n 
 d x$ which can be evaluated~\cite{Oloa2016} to $a_1=-(1+\gamma_{\text{Euler}})/( 2\pi) = -0.25102\ldots$, $a_2=\left (
 1+ \gamma_{\text{Euler}}+\gamma_{\text{Euler}}^2/2 \right)/(2 \pi^2) -1/48 =0.067508\ldots$, yielding $a_2-a_1^2=-1/48+(2 \pi)^{-2}$.}
\end{table*}


\end{document}